\documentstyle[12pt]{article}
\begin{document}
\begin{flushright}DOE-ER/40682-33\\
      CMU--HEP94--08\\
      PITT--94--04\\
      February,1994
\end{flushright}
\vspace{0.3in}
\centerline{{\large \bf A Note on Thermal Activation}}
\vspace{0.2in}
\centerline{Daniel Boyanovsky,$^{(a)}$\ \ Richard Holman,$^{(b)}$}
\vspace{0.2in}
\centerline{Da-Shin Lee,$^{(a)}$\ \  Jo\~{a}o P. Silva,$^{(b)}$\ \
Anupam Singh $^{(b)}$}
\vspace{0.3in}
\centerline{{\it $^{(a)}$Department of Physics and Astronomy,}}
\centerline{{\it University of Pittsburgh, Pittsburgh PA~~15260}}
\centerline{{\it $^{(b)}$Physics Department, Carnegie Mellon University,
Pittsburgh, PA~~ 15213}}

\vspace{.3in}
\baselineskip=24pt
\centerline{\bf Abstract}
\begin{quotation}
Thermal activation is mediated by field configurations that correspond to
saddle points of the energy functional. The rate of probability
flow  along the unstable
functional  directions, i.e the activation rate, is usually obtained from  the
imaginary part of a suitable analytic continuation of the  equilibrium free
energy. In this note we provide a real-time, non-equilibrium interpretation of
this imaginary part which is analogous to the real-time interpretation of
the imaginary part of the one-loop
effective potential in theories with symmetry breaking. We argue that in
situations in which the system is strongly out of equilibrium the rate will be
time dependent, and illustrate this with an example.
\end{quotation}

\newpage

The analysis of thermal activation has always been an important and interesting
topic in different areas in physics. Recently, however, this subject has
assumed even greater importance due to the observation that there are field

configurations in the standard model (so-called
sphalerons\cite{sphaleron,mottolawipf,mclerran}) that mediate  baryon number
violating transitions which are {\it unsuppressed} at high  temperature.
Needless to say, this has important implications for the evolution of  baryon
number in the early universe\cite{baryogenesis}. Another reason why there  has
been a rekindling of interest in the evolution of metastable states in  the
early universe is due to the development of viable models of inflation
(extended inflation\cite{lastein}) that go back to Guth's\cite{guth} idea of
ending the inflationary era via false vacuum decay. In short, there are good
reasons for making sure that the decay of metastable states at finite
temperature  is understood from as many different perspectives as possible.

The motivations for this note are twofold. First, we offer a real-time
interpretation of the standard equilibrium calculation of the activation rate
based on the imaginary part of an analytically continued free energy. Here we
draw an analogy with the case of the imaginary part of the effective potential
in theories with spontaneous symmetry breaking as studied by Weinberg and
Wu\cite{weinbergwu}. Second, we will make the point that in an  out of
equilibrium situation, different initial states can make for significant
changes in the activation rate. We show by an explicit example that in an out
of equilibrium situation the activation rate {\it must} be time dependent.

Two of the seminal works on this topic are those by Langer\cite{langer} and
Affleck\cite{affleck}. It will be instructive to review the concepts and
assumptions involved in these calculations, since our results will provide a
different interpretation of these standard results, as well as an extension of
them.

Langer\cite{langer,borkovec} develops a Fokker-Planck type equation for  the
probability of finding the system in a given configuration at time $t$. There
are two terms in this Fokker-Planck equation.
The first term is deterministic in nature, corresponding
to the Liouville time evolution in phase space. The other is a stochastic term
and represents the diffusion or dissipation arising from the coupling to
a heat bath.  The probability density in phase space obeys a continuity
equation and the associated current gives the flow of probability. In a
metastable situation, there is a particular configuration that corresponds to a
saddle point of the energy functional, and probability flows along the unstable
direction from the metastable phase towards the stable phase.
The activation (decay) rate is obtained as the total current passing over the
saddle point along the unstable direction divided by the initial population of
the metastable state\cite{borkovec}.

There are two very important assumptions in Langer's original work. The first
is that a steady state solution is reached in which a source in the metastable
state continually feeds probability (or particles) and a sink in the stable
phase removes probability at a steady rate. Secondly, the metastable state is
supposed to be in local thermodynamic equilibrium. This assumption
implies that there is a wide separation between the equilibration (relaxation)
time scale and the time it takes for the state to lose a significant amount of
probability. This in turn translates into the statement that there is fairly
strong dissipation\cite{borkovec}.  Under these assumptions, there is no
memory of the initial preparation of the metastable state.

Similar assumptions are implicit in Affleck's calculation\cite{affleck}. The
rate is defined as the Boltzmann average of the decay rates of the quantum
states that, for energies less than the barrier height, are standing waves in
the metastable well. For energies higher than the barrier height, these are
waves incident from the left, reflected and transmitted at the barrier. The
rate is then calculated as:

\begin{equation}
\Gamma = Z_0^{-1} \int_{0}^{\infty}\ dE \rho(E) \Gamma(E) \exp(-\beta\ E),
\end{equation}
where $\rho(E)$ is the density of states at energy $E$, $\Gamma(E)$ the decay
rate for states with energy $E$, and $\beta = 1\slash T$. The rate is
normalized using the partition function $Z_0$ of a harmonic oscillator centered
at the metastable state. As in Langer's calculation, this corresponds to a
steady flow of probability across the barrier.

Both of these calculations (and those that have built on them) assume that the
state under consideration is one in which probability is being fed continously
into the metastable well in order to replenish the probability that is flowing
out and over the activation barrier, thus ensuring a steady state.
Alternatively, the assumption is that the system is in quasi-equilibrium and
``slowly'' leaking probability on times scales much longer than typical scales
of the system. However, this implies that the system is looked at during times
for which all transient effects arising from the initial preparation have died
out and that the probability inside the metastable well has not been depleted
appreciably. In particular the underlying assumption\cite{borkovec} is that the
equilibration time inside the metastable well is much shorter than the inverse
rate of depletion of the population in the well.

These authors showed that under these (implicit) assumptions, the rate obtained
is the same as that obtained from the imaginary part of the analytically
continued free energy:

\begin{equation}
\Gamma= \frac{\Omega}{\pi T}\ {\rm Im} F \label{rateimfree}
\end{equation}
where $\Omega$ is the unstable frequency at the top of the barrier.

This description offers a fairly reliable method of calculation in a wide
variety of experimental situations in {\it macroscopic systems}\cite{borkovec},
in which there is a strong coupling to an environment (heat bath) and only a
few degrees of freedom are relevant.

However, the setting we have in mind is that of a {\em weakly} coupled
quantum field theory, in which the separation between system and bath is
somewhat blurred. In particular, at the onset of a first order phase
transition,
several time scales are relevant such as the equilibration rate, expansion
rate (in a cosmological situation) and the activation rate.

Given that this calculation of the rate involves  assumptions which in any
typical quantum field theory will be extremely hard to justify at the
microscopic level without invoking some phenomenological approximation, we want
to investigate an alternative {\em real-time} description of the process of
thermal activation.  We also envisage situations in which the system {\it may}
be strongly out of equilibrium, and in which some of the above assumptions are
not likely to apply.

To illustrate our discussion with a definite example, consider the case of a
scalar field theory  described by a potential with a local metastable minimum
and a global, stable  minimum. In this case, there is a static spherically
symmetric solution to the classical field equations that resembles a droplet of
the stable phase inmersed in a sea of the metastable phase\cite{langer,linde}.
This solution is characterized by the following  collective coordinates:
translational (``zero modes'') corresponding to the position of the droplet and
another that represents the radius of this droplet (although technically
speaking this latter does not correspond to a ``zero mode''). The static
solution to the field equation corresponds to a particular value of the radius
of the droplet, the critical radius, at which the surface energy is balanced by
the volume energy.

Fluctuations around this particular solution are characterized by zero
frequency modes (corresponding to translational invariance), a negative
eigenvalue corresponding to small fluctuations of the radius of the droplet,
and a spectrum of positive frequencies. Thus this configuration corresponds to
a saddle point in functional space. The unstable mode describes the instability
towards collapse or growth of the droplet for radii smaller or larger than the
critical radius, respectively, when either the surface energy or the volume
energy start to dominate.

Quantization around the classical solution corresponds to treating  the
position of the droplet  as a collective coordinate and
expanding\cite{christ,rajaraman,tomboulis,creutz,mattis}

\begin{equation}
\Phi(\vec{r},t)  =  \Phi_{cl}(\vec{r}-\vec{r}_o(t),R=R^*)+
\sum_l Q_l(t)f_l(\vec{r}-\vec{r}_o(t),R=R^*) \label{fiquant},
\end{equation}
where $\vec{r}_o$ is the position of the droplet, $R$ is the radius and $R^*$
is the critical radius for which the energy functional has an extremum. The
mode functions $f_l(\vec{r})$ are the eigenfunctions of the fluctuation
operator around the {\it static} droplet configuration and are chosen to be
orthogonal to the ``zero modes''\cite{mattis}. There is a coordinate $Q_u$
associated with an unstable mode with mode function $f_u(\vec{r}) \approx
d\Phi_{cl}/dR|_{R=R^*}$ and negative frequency $-\Omega^2$.
To lowest order in the
semiclassical expansion, the Hamiltonian operator becomes

\begin{equation}
H= E_{R^*}+\frac{\vec{P}^2}{2M}+\frac{P_u^2}{2}-\frac{1}{2}\Omega^2
Q_u^2 + \sum_{l'}\left[\frac{P_{l'}^2}{2}+\frac{1}{2}
\omega^2_{l'}Q_{l'}^2\right],
\label{semiclasham}
\end{equation}
with $E_{R^*}$ being the energy of the critical droplet, $M$ is
given by the normalization of the zero modes and $\omega_{l'}$ are the
stable frequencies.
To this order in the
semiclassical approximation, $\vec{P}$ is the total momentum of the
system. When the radius of the droplet is much larger than the correlation
length (thin wall approximation)

\begin{equation}
E(R) \approx 4\pi R^2 \sigma- \frac{4\pi}{3} R^3 \Delta{\cal{E}}
\end{equation}
with $\sigma$ being the surface tension given by the gradient terms and
${\Delta\cal{E}}$ the energy (or free energy) difference between the globally
stable and the metastable lower energy (or free energy) states for the
homogeneous configuration.
In this case the radius of the critical droplet is given by
\begin{equation}
R^* \approx \frac{2\sigma}{\Delta{\cal{E}}}
\end{equation}

In a saddle point approximation of the {\it equilibrium} partition function,
only two extremum configurations are usually considered. The first
corresponds to the metastable minimimun, while the second corresponds to the
droplet configuration (saddle point). The partition function is thus written as
the sum of these two independent terms.

Formally, the partition function evaluated at the saddle point corresponding to
the droplet does not exist because of the presence of the unstable (negative
frequency squared) oscillator.
However, it may be defined via an analytic continuation, in which case it
attains an imaginary part which in three spatial dimensions is given
by:

\begin{equation}
Z_{an} = \pm i V \left[\frac{MT}{2\pi}\right]^{3/2}
\frac{\exp[-\frac{E(R^*)}{T}]}{2\sin(\Omega/2T)} \prod_{l'}
\frac{1}{2\sinh(\omega_{l'}/2T)},
\label{partfunc}
\end{equation}
where the $\pm$ arises from the direction of the analytic continuation. The
volume factor $V$ and the $T^{3/2}$ factor arise from the ``free particle''
zero mode, the $1/\sin(\Omega/2T)$ from the unstable mode and the product over
$l'$ is only for the stable modes.
Writing the partition function $Z$ as:

\begin{equation}
Z=Z_o+Z_{an}=Z_o\left[1+\frac{Z_{an}}{Z_o}\right]\approx
Z_o\exp\left[\frac{Z_{an}}{Z_o}\right] \label{imfreener}
\end{equation}
with $Z_o$ the partition function of independent harmonic oscillators
at the metastable minimum of the potential, choosing the negative
sign for the analytic continuation (so as to obtain decaying exponentials),
and using the identification (\ref{rateimfree}) finally leads to the
usual formula for the rate per unit volume:

\begin{equation}
\frac{\Gamma}{V} =
\left[\frac{MT}{2\pi}\right]^{3/2}
\frac{\Omega \exp[-\frac{E(R^*)}{T}]}{2 \pi \sin(\Omega/2T)}\frac{\prod_{k}
2\sinh(\omega^o_{k}/2T)}{\prod_{l'} 2\sinh(\omega_{l'}/2T)}.
\label{usualrate}
\end{equation}
Here $\omega^o_{k}$ are the frequencies at the metastable minimum and
$\omega_{l'}$ are the {\it stable}  frequencies in the perpendicular
directions at the saddle point.

In functional space, the direction of steepest descent at the saddle point
corresponds (locally) to the unstable coordinate $Q_{u}$. For the purposes of
analyzing the decay of a metastable state and activation over the barrier (near
the top of the barrier), only this quantum mechanical coordinate is important.
Thus along this unstable direction, the problem becomes effectively that of a
quantum mechanical inverted oscillator near the top of the barrier. This
approximation implies that the temperature is such that $\omega_k \ll T <
E(R^*)$, where $\omega_k$  are the typical frequencies at the metastable well.

Another situation in which the free energy (or ground state energy) acquires an
imaginary part is in the calculation of the one-loop effective potential for
homogeneous configurations either at zero or finite temperature in the case
when the tree level potential $V(\Phi)$ has a double well
structure\cite{weinbergwu}.  In this case the fluctuations of wave vectors
$\vec{k}^2 < V''(\Phi)$ are unstable since they see an inverted harmonic
oscillator, just like the unstable coordinate above. These modes are treated
via an analytic continuation similar to the one performed above, giving rise to
an imaginary part for the one loop effective potential  (or free energy) for
homogeneous configurations\cite{boyvega}.

Weinberg and Wu\cite{weinbergwu} argued that this imaginary part of
the effective potential can be interpreted as the decay rate of a
gaussian state peaked at the ``top of the inverted oscillator'' for these
unstable modes. The rate of spread of this gaussian wave packet
was identified with the imaginary part of the one-loop effective
potential.

Further analysis\cite{boyvega} shows that this rate also determines the rate of
growth of fluctuations which is determined by the width of the gaussian state
initially prepared to be centered at the top of the inverted parabola at the
{\it maximum} of the tree level potential. This analysis was also carried out
at finite temperature, in which case the analytically continued free energy has
an imaginary part\cite{boyvega}. These unstable modes are responsible for the
early stage dynamics of a second order phase transition and for the
non-equilibrium evolution during the phase transition. The regions in which the
field becomes correlated grow in time as the system becomes more and more
correlated. Initially the unstable modes grow exponentially but eventually the
non-linearities set in, slowing down the rate of growth. Finally, at a later
time the growth stops. The system is now described as having large correlated
regions in which the field inside is close to one of the minima of the
(effective) potential. This is the mechanism of ``spinodal
decomposition''\cite{boyvega}. Thus the imaginary part of the
one-loop effective potential (or alternatively at finite temperature the free
energy) disguises a time dependent situation. Trying to study this time
dependent situation via an equilibrium description leads to treating the
unstable modes via an analytic continuation and an imaginary part.

The physical difference between the mechanism of spinodal decomposition and
thermal activation is that in the former there are no free energy
barriers, and small amplitude long wavelength fluctuations become
unstable and grow. By contrast, in thermal activation, there is a free energy
barrier to be overcome. Thus, a large amplitude configuration must be present
in the bath with a radius that is bigger than the critical radius.
Once this configuration is created, however, it is the instability
towards growth of this bubble that drives the phase
transition. The probability of finding these configurations in the
bath is determined by the Arrhenius-Kramer exponential factor explicit
in $Z_{an}$ above. When a critical bubble is formed, the system is
found at the top of the barrier, that is at the saddle point.
In the gaussian approximation, the situation corresponds to a gaussian
density matrix (the thermal equivalent of the gaussian wave packet
studied by Weinberg and Wu) centered at the top of the barrier, and
for the unstable coordinate corresponds to an inverted harmonic
oscillator.
{}From this point onwards, the analysis of the evolution, is then similar to
that
of Weinberg and Wu\cite{weinbergwu} and Boyanovsky and de Vega\cite{boyvega},
with the difference that the initial state is a thermal density matrix rather
than a wavepacket.

We see then that the imaginary parts of the free energy and/or the
effective potential usually found in the calculation of the
decay rates are a signal of non-equilibrium, real-time evolution.

Thus motivated by the situation with the one-loop effective potential we
 now present a real-time evolution of an initial
state that corresponds to a one-dimensional quantum mechanical system
describing the unstable coordinate $Q_u$. The reduction of the
multidimensional problem to one collective coordinate is justified
because this unstable coordinate determines (in a neighborhood of the saddle
point) the direction of steepest descent in functional space.
This is the direction along which  probability will flow.

The procedure is straightforward. Start with an initial density matrix
$\rho(t=0)$ and then evolve it in time via either the Liouville equation:

\begin{equation}
i \hbar \frac{\partial \rho(t)}{\partial t} = [H,\rho(t)],
\label{liouville}
\end{equation}
or via the solution to this equation:

\begin{equation}
\rho(t) = \exp(-\frac{i}{\hbar} H t)\rho(0)\exp(\frac{i}{\hbar} H t).
\end{equation}
Here $H$ is the Hamiltonian for the unstable coordinate
\begin{equation}
H_u = \frac{P_u^2}{2}-\frac{1}{2}\Omega^2
Q_u^2, \label{hamunst}
\end{equation}

Given the density matrix as a function of time, we can look at its position
space representation $\rho(Q_u,Q'_u;t)\equiv
\langle Q_u|\rho(t)|Q'_u \rangle$. The
current along the unstable direction is then found via:

\begin{equation}
J(Q_u,t) = \frac{\hbar}{2 i}(\frac{\partial}{\partial Q_u}-
\frac{\partial}{\partial Q'_u})
\rho(Q_u,Q'_u;t)|_{Q_u=Q'_u}\label{current}.
\end{equation}
Evaluating this current at the saddle point will then give us the transition
rate (or
activation rate) over the barrier.

Generalizing the analysis of Weinberg and Wu\cite{weinbergwu} and Boyanovsky
et. al.\cite{boyvega},
we will {\it assume} that the initial density matrix $\rho(t=0)$
 (only for the
unstable coordinate) is that of an {\it upright} harmonic oscillator
of frequency $\omega$ in thermal equilibrium at temperature $T=1/\beta$
given by\cite{feynman}:

\begin{equation}
\rho(Q_u,Q'_u;t=0)  =  {\cal{N}}(0) \exp\left\{-\frac{\omega}
{2\hbar \sinh(\beta \hbar \omega)}
\left[(Q_u^2+Q^{'2}_u)\cosh(\beta \hbar \omega)-2Q_uQ'_u\right]\right\}
 \label{initialdensmat}
\end{equation}
\begin{equation}
{\cal{N}}(0)            =
\sqrt{\frac{\omega}{2\pi\hbar\sinh(\beta \hbar \omega)}} \nonumber
 \end{equation}

The coordinate space expression for the time evolved density matrix is, for
$t>0$:
\begin{equation} \rho(Q_u,Q'_u;t) = \int\ dy \ dy' \langle
Q_u|\exp(-\frac{i}{\hbar}H_{u} t)|y\rangle \rho(y,y';t=0) \langle
y'|\exp(\frac{i}{\hbar}H_{u} t)|Q'_u\rangle \label{density} \end{equation}

The propagators $\langle x|\exp(\pm \frac{i}{\hbar}H_{u} t)|y\rangle$ are easy
to evaluate by analytically continuing the propagator for a standard harmonic
oscillator with real frequency\cite{barton}:

\begin{eqnarray}
\langle x|\exp(\pm \frac{i}{\hbar}H_{q} t)|y\rangle
         & = & N(t)
\exp(\pm \frac{i}{2 \hbar} \frac{\Omega}{\sinh(\Omega t)}\left[(x^2+y^2)
\cosh(\Omega t)-2xy\right]) \label{props} \\
N(t)     & = & (\frac{\pm \Omega}{2 \pi i \hbar \sinh(\Omega t)})^{1/2}.
\nonumber
\end{eqnarray}

It can be easily checked that these propagators are solutions to the evolution
equation with the proper boundary conditions. We can now compute the density
matrix as a function of time, as well as the current. We have verified that the
resulting density matrix is a solution of the Liouville equation
(\ref{liouville}) with the initial boundary condition given by
(\ref{initialdensmat}),  thus confirming that the analytically continued
propagators give the correct answer.

Rather than write down the density matrix, we consider the probability
density $p(Q_u,t) \equiv \rho(Q_u, Q_u;t)$:

\begin{eqnarray}
p(Q_u, t)  & = &   {\cal{N}}(t) \exp(-Q^2_u\slash {2 \sigma (t)^2})
\label{probdens}
\\
\sigma (t)         & = & \sqrt{\frac{\hbar}
{2\omega \tanh(\beta \hbar \omega/2)}}
\left[\cosh^{2}(\Omega t) +
 \frac{\omega^2}{\Omega^2}\sinh^{2}(\Omega t)\right]^{\frac{1}{2}}
\nonumber \\
{\cal{N}}(t)       & = & {\cal{N}}(0)
 \left[\frac{\sigma(0)}{\sigma(t)}\right] \nonumber
\end{eqnarray}

Thus, at long times, the probability to find the density matrix of the
unstable coordinate centered at the saddle point is

\begin{equation}
p(0,t) \approx \frac{{\cal{N}}(0)}
{[1+{\omega^2}\slash{\Omega^2}]^{\frac{1}{2}}}\exp[-\Omega t]
\label{psaddle}
\end{equation}
At zero temperature this is similar to the result obtained by
Weinberg and Wu\cite{weinbergwu} for a particular unstable mode by
identifying $\Omega$ with the unstable frequency for that particular
mode.

The probability current (\ref{current}) evaluated at the saddle is
zero because the density matrix spreads symmetrically around the
saddle point along the unstable direction. However, we may still
define a rate as

\begin{equation}
\Gamma_u = -\frac{d\ln[p(0,t)]}{dt} \approx \Omega \label{ratedef}.
\end{equation}
This expression will be valid for $t \gg \Omega^{-1}$. Integrating this rate in
all functional directions perpendicular  to $Q_u$ to obtain the total
probability at the saddle point and dividing by the partition function for the
harmonic oscillators at the metastable well, we finally obtain an alternative
definition for the rate per unit volume:

\begin{equation}
\frac{\Gamma_u}{V} =  \Omega
 \left[\frac{MT}{2\pi}\right]^{3/2}
\exp[-\frac{E(R^*)}{T}]
\left[\frac{\prod_{k} 2 \sinh(\omega^o_{k}/2T)}
{ \prod_{l'} 2 \sinh(\omega_{l'}/2T)} \right] \label{newrate}
\end{equation}
which differs from (\ref{usualrate}) by the prefactor.

Each contribution to the rate above has a very simple and clear
 real-time interpretation. The situation considered corresponds to
the system being described as a gaussian density matrix centered at
the saddle point. Along the unstable direction the density matrix
(similarly to a wave packet) spreads at a rate determined by the
unstable frequency $\Omega$. Thus the rate (\ref{newrate}) corresponds
to the total rate of spread of the initial density matrix (integrated
over all the perpendicular directions) divided by the total probability
in the metastable well. The underlying assumption here is that the
system has reached the saddle point and remains there in
quasi-equilibrium
and that the depletion of probability at the saddle is a consequence of the
spreading of the density matrix, although the statistical
average of the collective coordinate (radius
of the droplet) remains at the top of the barrier.

We may draw the following conclusion from this analysis. The
calculation of the rate via the imaginary part of the free energy,
which in Langer's work is derived from a steady state assumption,
corresponds to the assumption that the statistical density matrix is
centered at the saddle point, and that the expectation value of the radius of
droplet is the critical radius.
Hence, just as in the
case of the  one-loop effective potential, the imaginary part of
the free energy is disguising an intrinsically time dependent situation
and it will determine the rate of growth of fluctuations (width of the gaussian
density matrix)\cite{weinbergwu,boyvega}.
This can be easily understood from the fact that
the fluctuation in the unstable coordinate is given by

\begin{equation}
\langle Q^2_u(t) \rangle
\propto \sigma^2(t) \label{fluctuation}
\end{equation}
We can apply the same analysis for a very different situation, that
of the system initially away from the saddle and for which a steady
state assumption is not applicable. Thus we now consider activation as
an {\it initial condition} problem and look at the situation in which
the average radius of droplets is {\it smaller} than the critical.
This corresponds to an initial situation far from equilibrium and not
in the steady state. Such a condition corresponds to the situation
studied in a simplified $1+1$
dimensional  field theory model by Boyanovsky and de Carvalho\cite{dc}.

Again we will only concentrate on the unstable coordinate $Q_u$, and
consider the case in which the initial density matrix for this collective
coordinate is that of an upright harmonic oscillator of
frequency $\omega$ that is displaced a distance $Q_{uo}<0$
from the origin at the initial time $t=0$ and with zero average of the
momentum conjugate to this coordinate:
\begin{eqnarray}
\rho(Q_u,Q'_u;t=0)  & = & {\cal{N}}(0) \exp\left\{-\frac{\omega}
{2\hbar \sinh(\beta \hbar \omega)}
\left[((Q_u^2-Q_{uo}^2)+
(Q^{'2}_u-Q_{uo}^2))\cosh(\beta \hbar \omega) \right. \right.
 \nonumber \\
                    &   & \left. \left. -2(Q_u-Q_{uo})
(Q'_u-Q_{uo})\right]\right\}
 \label{initialshiftdensmat}
 \end{eqnarray}
with the same normalization factor as in (\ref{initialdensmat}).

We now solve the Liouville equation along the same steps used
in the previous case using the real-time propagators. We obtain
the probability as a function of time $p(Q_u;t)=\rho(Q_u,Q_u;t)$
as:
\begin{equation}
p(Q_u, t)   =    {\cal{N}}(t) \exp\left[-(Q_u-Q_{uo}\cosh(\Omega t))^2
\slash {2 \sigma (t)^2})\right]\label{probdensshif}
\end{equation}
with $\sigma(t)$ and ${\cal{N}}(t)$ given by (\ref{probdens}).  The current
(\ref{current}) along the unstable direction  evaluated
 at the saddle point  ($Q_u=0$) is found to be

\begin{equation}
J_u(Q_u=0;t) = \frac{\omega^2}{\Omega}{\cal{N}}(0)
|Q_{uo}| A(t)
\exp\left[-\tanh(\beta \hbar \omega/2) B(t)\right] \label{rate}
\end{equation}
with
\begin{eqnarray}
A(t) & = &  \frac{\sinh(\Omega t)}{\left[\cosh^2 (\Omega t) +
{\omega^2}\slash{\Omega^2} \sinh^2 (\Omega t)\right]^{3/2}}\label{Aoft} \\
B(t) & = & \frac{\omega
Q^2_{uo}}{\hbar}
\left[1 + \frac{\omega^2}{\Omega^2}\tanh^2 (\Omega t)\right]^{-1}
\label{Boft}
\end{eqnarray}

The current is shown in figure 1. The probability evaluated at the
saddle point is

\begin{equation}
p(Q_u=0, t)   =    {\cal{N}}(t) \exp\left[-(Q_{uo}\cosh(\Omega t))^2
\slash {2 \sigma (t)^2}\right]\label{probdensshif2}
\end{equation}
with $\sigma(t)$ and ${\cal{N}}(t)$ given by (\ref{probdens}).
At times $t \gg \Omega ^{-1}$, the rate

\[-\frac{d \ln[p(Q_u=0,t)]}{dt} \approx \Omega \]
coincides with (\ref{ratedef}) and only reflects the spread of the
initial density matrix. One can use  this rate as an alternative
definition in real time, leading again to the result (\ref{newrate})
after integrating along all the perpendicular directions in functional
space.

The depletion of probability at the saddle clearly reflects two physical
effects. The first is the spread of the density matrix determined by
$\sigma(t)$, while the second is the ``rolling'' of the unstable coordinate
down
the inverted harmonic oscillator potential. The same features give the total
current at the saddle. However, whereas the coordinate rolls down with a time
dependence $Q_{uo}\cosh(\Omega t)$, the spread of the density matrix increases
much faster and is given by $\sigma(t)$. The two effects thus combine to give a
{\it positive} current along the saddle point. We see then that despite the
fact that
bubbles with radii {\it smaller} than the critical radius collapse, they still
contribute a positive current over the barrier. The {\em total} current along
the unstable direction, evaluated at the saddle point is obtained by
integrating the current (\ref{rate}) along all the perpendicular
directions\cite{langer}, and restoring the exponential Arrhenius factor. It is
normalized by dividing by the partition function of the metastable well. This
procedure then yields:

\begin{equation}
{\cal{J}}_{T,u} = J_u(Q_u=0;t)
 \left[\frac{MT}{2\pi}\right]^{3/2}
\exp[-\frac{E(R^*)}{T}] \frac{\prod_{k} 2\sinh(\omega^o_{k}/2T)}{\prod_{l'}
2\sinh(\omega_{l'}/2T)} \label{newcurr}
\end{equation}
where the factors were obtained as in (\ref{newrate}) and $J_u(Q_u=0;t)$ is
given by (\ref{rate}).

There are two questions that require an answer at this point: i)
which one is a suitable definition of the rate, ii) what is the range of
validity of the approximations involved in obtaining the results
presented above. We answer these in turn:

i) We have seen that the definition of the rate given by (\ref{newrate})
coincides (up to the prefactor) with the oft quoted result in the literature
given by equation (\ref{rateimfree}). This definition corresponds to describing
the process of activation by assuming that the system is described by a
gaussian density matrix in terms of the small fluctuations around the critical
droplet, and that the activation rate is given by the rate of spread of the
probability for a critical droplet integrated along all perpendicular
directions in functional space. Although this seems a sensible definition of
the rate, it is {\it not} the definition of the rate of change of probability
for finding the system localized in the metastable state. This latter
definition corresponds to calculating the total probability current at the
saddle point as is obtained either from quantum mechanics or via the
Fokker-Planck description\cite{langer,borkovec} and corresponds to the current
(\ref{newcurr}) calculated above. The steady state assumption implicit in
Langer's treatment thus leads to the identification of the two. However, we
expect that in a strongly out of equilibrium situation, for which a steady
state assumption is not valid, or during times for which the details of the
initial state are important, the correct definition of the rate, that is, the
probability current at the saddle, will yield a time dependent rate which will
be sensitive to the initial conditions. In the case we have described above,
this corresponds to sensitivity to the initial value of the coordinate $Q_{u}$.

\noindent{ii)} There are several approximations invoked in this (and most
other) analysis. First, the quadratic approximation around the critical droplet
configuration. This approximation will be justified insofar as the higher order
terms in the expansion (cubic and quartic in terms of the coordinates $Q_l$)
can be neglected. Given that the imaginary part of the free energy describes
a time dependent process in which the width of the probability distribution
grows and therefore the fluctuations are growing, this approximation will break
down at some time that will be given by the details of the particular problem,
such as couplings and structure factors of the stable mode functions
$f_l(\vec{r})$. In order to identify the rate (\ref{newrate}) we had to assume
that $t \gg \Omega^{-1}$, for small $\Omega$ (as is expected in the thin-wall
approximation). This approximation may imply times longer than the regime of
validity of the gaussian approximation, in which case the rate may depend on
time. A quantitative analysis of the validity of the approximations will
require a more detailed knowledge of the potentials, mode functions couplings
etc. and will have to be done for each particular problem individually.

We started this work by using examples such as first order inflationary
models
and baryon number violation via sphaleron mediated decays to motivate the
discussion on thermal activation.
It is our impression that there is certain amount of discomfort and
uneasy trust on the usual calculation of the activation rate via
the imaginary part of an analytically continued equilibrium free energy.
We have drawn the analogy between the case of thermal activation and the
case of the one-loop effective potential for theories with spontaneous
symmetry breaking at tree level. Both calculations involve gaussian
approximations around unstable configurations (inverted harmonic
oscillators) and both calculations obtain an imaginary part by
analytically continuing the result for upright harmonic oscillators.

In both cases, the imaginary part of the free energy is associated with the
growth of the width of an initially prepared gaussian state (density matrix)
and signals the growth of fluctuations as a consequence of the instabilities.

We have shown that a real-time description of this process leads to
a definition of the rate that, up to prefactors, is similar to that
obtained via the recipe using the imaginary part of the free energy.

We also argued that in general, non-steady state or strongly out of
equilibrium situations the {\it correct} definition of the
rate, as the total probability current flowing across the saddle
point along the unstable direction will lead to a rather different
result which may in general be time dependent. We illustrated this
situation with a non-equilibrium case with an initial situation in which
the mean value of the collective coordinate representing the radius
of the droplet has an expectation value which is smaller than the
critical radius.
Time dependent rates are common in activated events in condensed
phases\cite{woly} where transient behavior after an initial preparation is
observed (see also\cite{borkovec}).

Although we cannot make general claims as to what is the correct rate in a
given theory, our goal here was somewhat more restricted: to provide a
dynamical, real-time  interpretation of the imaginary part of the free energy
and at the same time investigate scenarios for which a calculation of the rate
based on this analytic continuation may not yield reliable results.

We are  currently involved
in calculations of decay rates in theories involving sphalerons using the
real-time formalism developed here\cite{abelianhiggs} as an initial condition
problem. Specifically, if the phase transition occurs via a quench in
such a way that the system does not have time to equilibrate around the
minima, thermal activation will occur strongly out of equilibrium and
dependence on the initial state and time dependent rates are expected.

The issue of thermalization in the wells requires a deeper understanding
of the time scales involved, in particular in inflationary scenarios.

While we do not yet have all the answers we need to fully understand what
changes the time dependence of the rate will bring, we may speculate. We have
considered thermal activation here rather than under the barrier tunnelling.
This makes the range of applicability of our calculation to inflationary models
somewhat suspect. The reason for this is that once inflation sets in, the
temperature of the heat bath will decrease rapidly, turning the problem into a
zero temperature one. Even in this case, however, we should expect the rate to
be time dependent. The basic change in our calculation is that the paths used
to compute the propagators required to evolve the density matrix in time will
be
different. Essentially, one must do a WKB approximation of the
propagators\cite{danme}.

The situation in extended inflationary models is somewhat trickier to assess,
since the nucleation rate is already time dependent in most of these models due
to the time evolution of the Jordan-Brans-Dicke field in them\cite{us}.

In the case of the sphaleron, the question of whether the sphaleron
interactions are in thermal equilibrium (which is crucial in terms of
determining whether a baryon asymmetry can be generated by these interactions),
becomes more difficult to assess due to the time dependence of the rate. One
could imagine that the rate of these interactions
decreased sufficiently quickly so as to allow them to drop out of local thermal
equilibrium thus allowing a net $B$ asymmetry to be generated. However, the
answer to these and other questions will only be found
once a deeper understanding of the real-time dynamics of the process of
thermal activation is understood in these systems.

\vspace{18pt}

\centerline{\bf Acknowledgements}
D.B. would like to thank David Jasnow, Walter Goldburg and Larry McLerran  for
useful discussions. D.B. and D-S.L were supported in part by NSF grant  $\#$
PHY-9302534 as well as a Mellon Pre-Doctoral Fellowship Award (D-S.L). R. H.,
J.S. and A.S. were supported in part by DOE grant $\#$ DE-FG02-91ER40682. J.S.
was also supported in part by the Portuguese JNICT under CIENCIA grant $\#$
BD/374/90-RM

\centerline{\bf Figure Captions}

\vspace{5mm}

Figure 1: The thermal activation rate $\Gamma(t)$ as a function of $t$ at fixed
temperature. Time is measured in units of $\Omega^{-1}$.


\begin{thebibliography}{99}

\bibitem{sphaleron} For a good review, see ``Anomalous Fermion Number
Non-Conservation'' by M.E. Shaposhnikov, CERN preprint CERN-TH.6304/91 (1992).

\bibitem{mottolawipf} E. Mottola and A. Wipf, Phys. Rev. D39, 588 (1989).

\bibitem{mclerran} P. Arnold and L. McLerran, Phys. Rev. D36, 581 (1987);
Phys. Rev D37, 1020 (1988); L. McLerran, E. Mottola and M. E. Shaposhnikov,
Phys. Rev. D43, 2027 (1991).

\bibitem{baryogenesis} See e.g. ``Non-GUT Baryogenesis'' by A. D. Dolgov
 Phys. Rep. {\bf 222}, 309 (1992).

\bibitem{lastein}D. La and P. J. Steinhardt,  Phys. Rev. Lett.{\bf 62}, 376
(1989).

\bibitem{guth} A.H. Guth, Phys. Rev. D{\bf 23} 347 (1981).

\bibitem{weinbergwu} E. Weinberg and Wu, Phys. Rev. D{\bf
36}, 2474 (1987)

\bibitem{langer}J.S. Langer,  Ann. Phys. (N.Y.){\bf 41} 108 (1967); {\em ibid}
{\bf 54} 258 (1969). Solids Far From Equilibrium ed. by C. Godreche (Cambridge
University Press, Cambridge, 1992), p. 297.



\bibitem{affleck}Ian Affleck,  Phys. Rev. Lett.{\bf 46} 388 (1982).

\bibitem{borkovec}  For a comprenhensive review  see:
P. Hanggi, P. Talkner and M. Borkovec, Rev. of Mod. Phys. 62, 251
(1990).


\bibitem{christ} N. H. Christ and T. D. Lee, Phys. Rev. D 12,
 1606 (1975).

\bibitem{rajaraman} R. Rajaraman, Solitons and Instantons, (North Holland,
Amsterdam, 1982).

\bibitem{tomboulis} E. Tomboulis, Phys. Rev. D 12, 1678 (1975).

\bibitem{creutz} M. Creutz, Phys. Rev. D 12, 3126 (1975).



\bibitem{mattis} There are some important aspects of
quantization around classical configurations in higher order that deserve more
attention. A thorough analysis is
given by N. Dorey and M. Mattis ``Soliton quantization and
internal symmetry'' (to appear in Phys. Rev. D (1994)).


\bibitem{boyvega} D. Boyanovsky and H. J. de Vega, Phys. Rev. D
{\bf 47}, 2343 (1993); D. Boyanovsky, D-S. Lee and A. Singh,
Phys. Rev. D{\bf 48}, 800 (1993). D. Boyanovsky, Phys. Rev. E{\bf 48},
767 (1993).

\bibitem{linde} A. Linde ``Particle Physics and Inflationary
Cosmology'' (Harwood Academic, 1990)


\bibitem{feynman} `` Statistical Mechanics'' by R.P. Feynman,
W.A. Benjamin Inc., Reading, Mass. (1972).

\bibitem{barton} G. Barton,  Ann. Phys. (N.Y.) {\bf 166} 322 (1986).


\bibitem{dc} D. Boyanovsky and C. Arag\~{a}o de Carvalho,
 Phys. Rev. D48, 5850 (1993).

\bibitem{woly} P. G. Wolynes, Phys. Rev. Lett.{\bf 47}, 968 (1981).

\bibitem{abelianhiggs}D. Boyanovsky, R. Holman, D-S. Lee, J.P. Silva and
A. Singh, (work in progress)(1994).

\bibitem{danme}D. Boyanovsky, R. Holman and R. Willey,  Nucl. Phys.{\bf B376},
599 (1992).



\bibitem{us} R. Holman, E.W. Kolb, S.L. Vadas and Y. Wang,  Phys. Lett.
{\bf 250B}, 24 (1990).
\end{thebibliography}
\end{document}